**Single Atomic Fe anchored Porous Carbon with Rich Graphitic Nitrogen as Electrocatalysts for Oxygen Reduction Reaction and Zn-Air Batteries**

Yuanyuan Zhang[a,#], Shuo Wang[a,#], Cheng-Zong Yuan[a,#], Kwan San Hui[b,*], Chenghao Chuang[c,*], Qinghua Zhang[d], Lin Gu[d,*] and Kwun Nam Hui[a,*]

[a] Joint Key Laboratory of the Ministry of Education, Institute of Applied Physics and Materials Engineering, University of Macau, Avenida da Universidade, Taipa, Macau SAR, China

[b] Engineering, Faculty of Science, University of East Anglia, Norwich, NR4 7TJ, United Kingdom

[c] Department of Physics, Tamkang University, Tamsui 251, New Taipei City, Taiwan

[d] Institute of Physics, Chinese Academy of Sciences, Beijing, 100190, China

[#] These authors contributed equally

[*] Corresponding author

Kwan San Hui, E-mail: k.hui@uea.ac.uk

Kwun Nam Hui, E-mail: bizhui@um.edu.mo

Chenghao Chuang, E-mail: chchuang@mail.tku.edu.tw

Lin Gu, E-mail: l.gu@iphy.ac.cn

**Abstract**

Zn-air battery (ZAB) has distinguished itself as new generation of energy storage device due to the high theoretical energy density and its performance relies on the oxygen reduction reaction (ORR) performance of the cathode catalysts. Single atomic Fe anchored N-doped carbon (Fe-N-

C) has emerged as a promising ORR electrocatalyst because of the maximum utilization of Fe atoms. However, to obtain high-rate and stable Fe-N-C remains challenging. A novel and facile approach to fabricate Fe-N-C catalyst (PC-Fe-50) with outstanding ORR performance superior to commercial platinum catalyst and iron phthalocyanine (FePc), is proposed here. When mixed with commercial OER catalyst ($RuO_2$) and employed as the air cathode in ZAB, a high energy density of 809 W h $kg^{-1}$, high power density of 128 mW $cm^{-2}$, and stable cycling rechargeable performance are obtained. By means of density functional theory calculations, we revealed that the abundant N dopants (7.47 at%) in carbon play significant roles on $FeN_x$ moieties with two most common configurations ($FeN_4C_{10}$, denoted as D1; $FeN_4C_{12}$, denoted as D2). The binding energies of ORR intermediates on Fe center are adjusted. By comparing the activity of possible structures and FePc molecule, we find the realistic active sites in PC-Fe-50 catalyst may be the D2 combining the adjacent N atoms, instead of D1, the widely recognized active structure in reported Fe-N-C catalysts.

## 1. Introduction

New generation of energy storage devices has been urgently called for under the background of global energy crisis. Rechargeable Zn-air battery (ZAB) has distinguished itself due to the high theoretical energy density (1086 W h $kg^{-1}$), low cost and environmental benefits[1]. The performance of ZAB relies heavily on the oxygen reduction reaction (ORR) performance of the cathode catalysts. To date, the commercial platinum (Pt) catalysts remains the most efficient electrocatalyst for ORR but the high cost and poor stability hinder its large-scale application.

Single atomic Fe anchored N-doped carbon (Fe-N-C)[2] has emerged as a promising substitute ORR catalyst wherein the $FeN_4$ moieties show superior intrinsic activity with the maximum Fe atoms efficiency. Many efforts have been made to enhance the performance of Fe-N-C to a comparative level with Pt catalyst by creating densely dispersed mental single atoms[2b, 2e, 3] and fabricating atomic dual metal pairs sites precisely[2f, 2g]. Despite the progress achieved by those work, it is still challenging to prepare Fe-N-C through facile and low-cost method.

The content and species of N doped in carbon was proved to be one of the key factors for the ORR performance[4]. It is reported that the pyridinic N enhances the onset potential [4a] and the graphite N determines the diffusion limiting current density ($J_L$) value of the ORR[4a] and the adjacent carbon atoms act as the intrinsic active sites[4b], while the pyrrolic N was considered ineffective to the ORR[4a].

Moreover, in SAC the N plays an important role beside anchoring the metal atoms and forming the $MN_x$ sites. Recently, Xia, et al. revealed the a synergistic effect catalyst showed between graphitic N and the neighboring Fe center, graphitic N increased the filling degree of d-orbitals of the Fe and the optimized the binding energies of ORR[2c].

In most cases the N content in carbon-based materials was no more than 3 at%[2c, 2f, 4b, 5]. Recently, employing g-$C_3N_4$ as precursor is an efficient way to improve the total N content in the final products (10.07 at%[6], 8 at%[7]). In this case the pyridinic N and pyrrolic N were dominant species with small amount of graphitic N (0.54 at%[6], 2 at%[7]). However, it is still challenging to synthesize Fe-N-C with rich N content.

In this work, we demonstrate a facile two-step solvothermal and pyrolysis strategy to fabricate a single atomic Fe anchored porous carbon (marked as PC-Fe-50) catalyst with abundant N dopants (7.74 at%), containing graphitic N (3.11 at%) and pyridinic N (2.21 at%) exist as the dominant species. After optimizing the Fe content (0.65 at% by XPS or ICP) and the pyrolysis temperature (900 °C), this novel catalyst exhibits an excellent ORR catalytic activity in $O_2$-saturated 0.1 M KOH, with high onset potential of 0.96 V, $J_L$ of 9.33 mA cm$^{-2}$, and robust durability, which are notably superior to the commercial 20% Pt/C catalyst and iron phthalocyanine (FePc). We analyze the ORR activity of possible $FeN_x$ sites of two most common configurations ($FeN_4C_{10}$, denoted as D1; $FeN_4C_{12}$, denoted as D2) by the density functional theory (DFT) calculation. D1 has been widely recognized as active sites in Fe-N-C while D2 is ineffective. But DFT shows the activity of D1 is close to that of FePc molecules. However, considering that the abundant N atoms may be located adjacent to $FeN_x$ sites in PC-Fe-50, the binding energies for ORR intermediates on both D1 and D2 are adjusted due to….(to be finished). D2 can thus be largely activated and become more effective than both D1 and FePc. Moreover, the ZAB based on PC-Fe-50 mixed with commercial OER catalyst $RuO_2$ as air cathode delivers high energy density of 809 W h kg$^{-1}$, high power density of 128 mW cm$^{-2}$, and stable rechargeable performance. This work not only pave a new way to fabricate novel Fe-N-C for high-performance ZAB but also for the first time propose the D2 combining the adjacent N atoms, as the realistic active sites.

## 2. Results and Discussion

The synthesis process of Fe-PC-50 is schematically illustrated in Figure 1a. Typically, the Fe anchored carbon precursor was synthesized from 6 g of urea, 3 g of citric acid and 50 mg of FePc via the solvothermal method in 50 mL of Dimethylformamide (DMF) under 160 °C for 6 h. The resulting product was annealed under the flowing $N_2$ at 900 °C for 1 h to remove the hydrophilic functional groups (-COOH, -OH, etc.)[8]. Additionally, the PC, PC-Fe-25, PC-Fe-50 and PC-Fe-75 catalysts were prepared under the same conditions with no FePc, 25 mg FePc and 75 mg FePc, respectively. The samples annealing under 800 °C and 1000 °C were denoted as PC-Fe-50-800 and PC-Fe-50-1000, respectively.

From the micromorphology showed by the scanning electron microscopy (SEM) images (Fig. S1c) and the transmission electron microscopy (TEM) images (Fig. 1b, c), the PC-Fe-50 presents porous structure stacked by carbon bulks with a quasi-spherical shape and a diameter around 50 nm, which is similar to PC (Fig. S1a, Fig. S2a, b) and the samples with different Fe content, PC-Fe-25 and PC-Fe-75 (Fig. S1b, d), confirming that Fe didn't change the basic morphology and structure of carbon matrix, possibly due to the low Fe content (0.9-2.57 wt% by the inductively coupled plasma-atomic emission spectrometry (ICP-AES), Table S1; 0.27-0.74 at% by X-ray photoelectron spectroscopic (XPS), Table S2). Furthermore, for the PC, PC-Fe-25, PC-Fe-50 and PC-Fe-75 samples, the Raman curves with the comparable value of D band to G band ratio ($I_D/I_G$ ≈ 1.03-1.04, Fig. S3a) prove that Fe didn't influence the disorder degree of carbon lattice, either. The PC-Fe-50-800 shares the similar morphology with PC-Fe-50 (Fig. S1e), while the carbon particles in PC-Fe-50-1000 agglomerated to form big blocks (marked with red circles in Fig. S1f). Meanwhile the $I_D/I_G$ value in Raman spectra for PC-Fe-50-800 and PC-Fe-50-1000 are 1.1 and 1.01, respectively (Fig. S3b), indicating that the higher annealing temperature results in lower level of disorder and higher level of graphitization of carbon lattice, vice versa, consistent with the previous research[9].

The high-angle annular dark-field scanning transmission electron microscopic (HAADF-STEM) image and energy dispersive X-ray spectroscopy (EDS) mappings present the uniform distribution of C, N, Fe, O elements (Fig. 1e-i), and no visible nanoparticle appears in the high-resolution TEM (HRTEM) image (Fig. 1c), indicating that the Fe exists as isolated single-atoms in the PC-Fe-50 catalyst. As shown in Fig. 1d, the isolated bright dots marked with red circles further confirm the existences of Fe single atoms by the aberration-corrected HAADF-STEM.

Meanwhile, the X-ray diffraction (XRD) patterns of PC, PC-Fe-25 and PC-Fe-50 show only typical graphite (002) and (100) peaks at 26 ° and 44 ° without peak belonging to any possible crystallized Fe species (Fig. S4a), indicating that the atomic dispersed Fe was anchored on carbon under the 900 °C annealing temperature. When the Fe content increased (75 mg FePc in raw materials), nanoparticles were observed in TEM image (Fig. S5a) and several minor peaks around 45° appear and can be assigned to the crystallized $Fe_3C$ (PDF#35-0772), which was reported to be less active than $FeN_x$ sites for ORR[2c, 10], indicating that the excessive Fe atoms aggregated and formed $Fe_3C$ nanoparticles. Similar peaks also are shown in the XRD curve of the PC-Fe-50-1000, suggesting when the annealing temperature increased to 1000 °C (Fig. S4b), Fe also tended to react with C and to produce $Fe_3C$ nanoparticles (Fig. S5b). Note that the (002) lattice plane with a d-spacing of 0.35 nm (Fig. 1c), corresponding to the regular graphitic carbon layer, which appeared frequently in HRTEM images of other kind of graphitic carbon materials[2c, 5c, 11], cannot be observed here. Only few short fringes are visible in this work (Fig. 1c), suggesting the short-range order structure of graphite micro-crystals and the lattice distortion possibly caused by abundant N dopants (6.9-7.9 at % measured by XPS, Table S2).

The surface area and pore structure of the as-synthesized samples are analyzed by the $N_2$ adsorption-desorption experiments and Brunauer-Emmett-Teller (BET) method (Fig. 2a-d). For all samples the curves display a similar III-type isotherm characteristic[12] of macro porous owing to the low surface area and high pore volume characteristics of the PC (93.2 $m^2$ $g^{-1}$ and 0.72 $cm^3$ $g^{-1}$). The PC-Fe-25 (87 $m^2$ $g^{-1}$ and 0.22 $cm^3$ $g^{-1}$), the PC-Fe-50 (63.8 $m^2$ $g^{-1}$ and 0.27 $cm^3$ $g^{-1}$) and the PC-Fe-75 (49 $m^2$ $g^{-1}$ and 0.25 $cm^3$ $g^{-1}$) exhibited a obviously decreased surface area and lower total pore volume than the PC matrix (Table S4). These can be explained by the nucleation and growth mechanism in the solvothermal process. For the solvothermal reaction of urea and citric acid without Fe, carbon particles were obviously formed through random and scattered nucleation from small molecules in DMF solution during the solvothermal process. For the precursor of PC-Fe-X (X = 25, 50, or 75), the samples with Fe, the reaction process is different. According to the theory of heterogeneous nucleation mechanism, the interactions with the formation of new phase nuclei are running in contact either with heterogeneities found in the generating phase, or with the surface[13]. It can thus be concluded that the carbon tended to grow around FePc molecules dispersed in solution. Hence, for the PC-Fe-25 and PC-Fe-50, carbon particles consisting structures of $FeN_x$ sites anchored in[14], or connecting graphene sheets[15], was formed (corresponding to two different structures, shown in Fig. 1a). Also, the decreased specific surface area value of PC-Fe-X samples than PC can be explained by the assemblies of carbon around FePc. Meanwhile, the metal single atoms (in PC-Fe-25 and PC-Fe-50) and nanoparticles (in PC-Fe-75) block some pores and connected channels in carbon support[2a, 10, 16] and decrease the pore volume significantly. The specific surface area also decreased with the annealing temperature (Fig. 2c, Table S4), indicating the carbon particles tend to be aggregated to form big bulks under high temperature, which agree with the SEM images of PC-Fe-50-1000 (Fig. S1f). The pore size distribution curves (Fig. 2b, d) reveal two peaks associated with two pore width ranges. The mesopores with size of 2~4 nm might be assigned to the distortion in carbon matrix. While the mesopores with size of 20~40 nm could be ascribed to space among the stacked carbon particles. Although the content of mesopores was decreased due to the introduce of Fe or the high annealing temperature, the remaining mesopores still play a vital role in creating more active sites in defects and facilitating the electrolyte/reactant diffusion during the ORR process[4b, 17].

The XPS measurements were carried out to probe the chemical species and configurations of elements in samples. In high-resolution N 1s spectra (Fig. 2e), the peaks located on 402.4, 401, 399.5, 398.2 eV can be indexed to the oxidized, graphitic, pyrrolic and pyridinic N, respectively[4a, 4b, 9, 18]. For PC-Fe-50, the peak at 398.9 eV can be assigned to Fe-N [2a, 19], proving the existence of Fe-N-C bonding. The content of graphitic, pyridinic and pyrrolic N are calculated to be 3.7, 2.44 and 0.63 at% by XPS, respectively (Table S3), and the overall N content in the PC is 7.9 at% (Table S2). Notably, the quantitative ratio of N and C atoms is close to 1:12, signifying that one N atom exists within two 6-C-rings averagely, suggesting a much higher doping content than most previous reported N-doped carbon catalysts (Table S5). As for the PC-Fe-50, the content of N, graphitic N, pyridinic N and pyrrolic N slightly decreased to 7.5, 3.11, 2.21 and 0.68 at%, respectively. Considering the Fe source, FePc, introducing N atoms along with Fe, it can be inferred that, to a small extent, the FePc molecules hindered the participation of the urea molecules in the synthesis of C atom rings during the solvothermal process, and the N content in PC-Fe-25 and PC-Fe-75 also followed this rule (Fig. S6, Table S2). The influence of pyrolysis temperature on N content can be obtained by comparing the N content of PC-Fe-50-800 (5.9 at%) and PC-Fe-50-1000 (4.6 at%) with that of PC-Fe-50. It can be concluded that either higher or lower pyrolysis temperature than 900°C didn't benefit for the N doping into C lattice. Specially, the content of pyridinic and pyrrolic N are only 0.8 and 0.2 at% in PC-Fe-50-1000, suggesting that these two N configurations are less thermolabile than graphitic N under high temperature[4b, 9].

As deconvoluted from the high-resolution Fe 2p spectra (Fig. S7), the two spin–orbit doublets at 709.9 and 723 eV, correspond to the Fe $2p^{3/2}$ and Fe $2p^{1/2}$ orbitals, with satellite peaks at 717.7 and 729.5 eV, respectively. While the peaks at 713.5 and 726.2 eV can be attributed to the Fe-N configuration, confirming the existence of FeN$_x$ moiety[2d, 11, 20]. X-ray absorption near-edge structures (XANES) was performed for compare the valence state of PC-Fe-50 and other standard reference samples. As shown in Fig. 2f, the XANES spectra reveal that PC-Fe-50 has a higher photon energy curve between FeO and Fe$_2$O$_3$, meaning that the stable valence state Fe in the PC-Fe-50 is between +2 and +3, consistent with the results of reported Fe SAC samples[2b, 14]. Meanwhile, as a standard reference sample, the FePc holds characteristic pre-edge peak at 7112-7116 eV due to a 1$s$-4p$_z$ square-planar configuration with high *D4h* shakedown transition characteristic for a symmetry, which was regarded as a fingerprint of FeN$_4$ square-planar structures[21], while the absence of a pre-edge peak in the PC-Fe-50 indicates a broken *D4h* symmetry and might due to a dioxygen from air were axially absorbed on the Fe center[14].

The ORR electrocatalytic activity of as-synthesized samples was investigated in O$_2$-saturated 0.1 M KOH solutions with RDE. The LSV of PC present a onset potential (defined by the potential at which the current density reaches 0.2 mA cm$^{-2}$ in this paper) of 0.83 V and a high J$_L$ of 7.4 mA cm$^{-2}$ (Fig. 3a), should be partially ascribed to the high graphitic N content of 3.7 at%, as the previous work proved that the graphitic N content determined the J$_L$[4a] possibly due to that the graphitic N actives the 3 closest carbon atoms[4b]. As for the PC in this work, both higher J$_L$ and higher graphitic N content compared to most N-doped carbon catalysts (Table S5) corroborate the above conclusion. The introduce of FeN$_x$ sites significantly improved the ORR activity of PC (Fig. 3a). Among the tested samples, the PC-Fe-50 shows a super outstanding ORR activity in term of onset potential of 0.96 V and half-wave potential of 0.78 V. Particularly its J$_L$ reaches to 9.33 mA cm$^{-2}$, higher than that of most catalysts ever reported in the same test conditions (Table S5). The PC-Fe-25 shows a less positive onset potential of 0.91 V and a lower J$_L$ of 8.51 mA cm$^{-2}$ due to the less dense FeN$_x$ sites. As for the PC-Fe-75, although it exhibits a slightly higher J$_L$ of 9.6 mA cm$^{-2}$, it is obviously less active in terms of the onset potential of 0.93 V and the half-wave potential of 0.75 V, should be attributed to the exist of less active Fe$_3$C nanoparticles[2a, 10]. The remarkable ORR activity of PC-Fe-50 is also revealed by the high double-layer capacitance (C$_{dl}$) value of 24.5 mF cm$^{-2}$ (Fig. 3b, Fig. S7). The effect of pyrolysis temperature also played a critical role on the activity of catalysts. Compared to PC-Fe-50, the catalyst pyrolyzed under lower temperature of 800 °C, even with a high C$_{dl}$ of 27.4 mF cm$^{-2}$, shows a poorer ORR performance, however, which resulted from the lower graphitization degree (higher $I_D/I_G$ value than other samples, Fig. S3b) and thus lower electrical conductivity of carbon[9]. While the poorer performance of PC-Fe-50-1000 and low C$_{dl}$ of 12 mF cm$^{-2}$ should be ascribed to the aggregation of carbon support and the formation of Fe$_3$C nanoparticles. Overall, PC-Fe-50 is proved to be the most efficient ORR catalysts among all the as-synthesized samples.

We evaluated the PC-Fe-50 catalyst together with the commercial 20% Pt/C catalyst and the raw FePc, which has also been employed as ORR catalyst[22] (Fig. 3e). Remarkably, PC-Fe-50 is close to Pt/C in term of onset potential but with a significantly higher J$_L$. Meanwhile, PC-Fe-50 exhibits much higher current response than FePc in the whole applied voltage rage (1.0~0.2 V). Most strikingly, the PC-Fe-50 demonstrates metal mass activity of 471.2, 242.4 and 33.7 A g$^{-1}$ at 0.7, 0.8 and 0.9 V, respectively, far exceeding that of Pt/C and FePc (Fig. 3f). As expected, the PC-Fe-50 also displays lower ring current, lower H$_2$O$_2$ yield and higher electron transfer number (3.9~3.98) than Pt/C (Fig. 3g, h), demonstrating a dominant direct 4-electron transfer pathway and high selectivity towards ORR. Furthermore, PC-Fe-50 exhibits the long-term stability with a higher current retention (97.6 %) than Pt/C (Fig. 3i). Also, there is only slightly decay in J$_L$ after 1000 continuous potential cycles (Fig. S8), confirming the excellent durability of PC-Fe-50.

**DFT**

Numerous studies based on the XANES and Mössbauer spectroscopy suggested that the single atomic $FeN_x$ sites on carbon mainly consist of $FeN_4$ moieties in Fe-N-C catalysts[2c, 14-15, 23]. We summarize the constitution of $FeN_4$ sites in reported Fe-N-C in Table S6. Obviously the $FeN_4C_{10}$ (generally denoted as D1) and $FeN_4C_{12}$ (D2) have been regarded as the two dominant components of $FeN_4$ sites. We therefore assume that the $FeN_x$ sites in PC-Fe-50 are mainly D1 and D2. From the point of geometry, D2 are expected to bridge graphene sheets or locate on the edges of micropores while D1 anchor in the plane of graphene sheets without causing much distortion (as shown in Fig. 1a)[14-15, 23a, 23b]. Both theoretical[23a, 24] and experimental[23d-f] study which have fully proved that D1 is highly active to ORR while D2 is relatively inactive and D1 has been commonly acknowledged to be responsible for ORR in Fe-N-C catalysts[2b, 2c, 3, 14, 25].

To identify the active sites and reveal the mechanism for ORR in PC-Fe-50, we firstly build atomic models (Fig. S11, 12) and calculated the free energy diagrams for ORR on D1, D2 structures and FePc molecule (Fig. 5a). The free energy diagram on D1 shows that when the overall reaction potential increased to 0.691 V, the fourth elementary step (*OH transfer to $OH^-$) reaches equilibrium and become a rate-determining state (RDS) with a corresponding computational overpotential for ORR ($\eta_{ORR}$) of 0.539 V. For D2, the RDS is the third elementary step (*O transfer to *OH) with the $\eta_{ORR}$ value of 0.718 V, which indicates an obvious low ORR activity compared to D1. This result is consistent with the previous research[23d-f, 24]. The free energy diagram of FePc was also calculated, the RDS is the third elementary step and the $\eta_{ORR}$ turned out to be 0.553 V, indicating that the activity of FePc is slightly lower than but very closed to D1. However, our experimental result shows the PC-Fe-50 possessed remarkably higher onset potential (0.96 V) than FePc (0.88 V, Fig. 3e). Moreover, the PC-Fe-50 demonstrates metal mass activity (normalized by Fe content) of 471.2, 242.4 and 33.7 A $g^{-1}$ at 0.7, 0.8 and 0.9 V, respectively, many times higher than those of FePc (less than 10 A $g^{-1}$, Fig. 3f). This experimental result indicates that the realistic sites in PC-Fe-50 should be far more active than FePc molecules, which is inconsistent with DFT results.

As described earlier in this paper, the N content in PC-Fe-50 is 7.47 at% and higher than in most reported N-doped carbon catalysts (Table S5). Suggesting the N atoms disperse uniformly on carbon substrate, there is a considerable probability (>50%) that at least one N atom locate adjacent to and have impact on the $FeN_4$ sites. We reconstructed several possible atomic structures in which the N atoms adjacent to D1, D2 sites were considered (as shown in Fig. S11) and DFT calculations were reperformed to evaluate their ORR activity (Fig. 5a, Fig. S11, Table S8). The D1-xN and D2-xN (x = 1, 2, 3, 4) represent D1 and D2 sites combining with x N atoms, respectively. Table S8 shows that the adsorption energies of intermediates, RDS and the $\eta_{ORR}$ of $FeN_4$ sites were significantly adjusted by adjacent N atoms. Among them, D1-2N, D1-3N, D1-4N show lower $\eta_{ORR}$ than D1 while D2-1N, D2-2N, D2-4N exhibit lower $\eta_{ORR}$ than D2. Specially, D1-4N and D2-2N turn out to be the most active structures with $\eta_{ORR}$ of 0.374 V and 0.319 V, respectively, which are obviously more active than FePc (Fig. 5b) and this is in good accordance with our experimental findings. However, considering that it should be more possible for two N atoms than four N atoms to locate adjacent to $FeN_4$ sites, D2-2N could be the realistic active site for ORR in PC-Fe-50.

To evaluate the catalytic properties of PC-Fe-50 in real applications, ZABs were assembled consisting of. As revealed in Fig. 4a, one single ZAB equipped with the PC-Fe-50+$RuO_2$ catalyst can provide enough voltage to power a white LED bulb (1.0 V). The battery with the PC-Fe-50+$RuO_2$ exhibits a voltage platform of 1.24 V at 5 mA $cm^{-2}$ (Fig. 4b, c), higher than that of Pt/C+RuO$_2$ (1.19 V). After normalizing to the weight of the dissipative Zn foil, a high specific capacity and energy density, 714 mA h g$^{-1}$ and 809 W h kg$^{-1}$, are achieved for PC-Fe-50+RuO$_2$, which are obviously higher than those of Pt/C+RuO$_2$ (626 mA h g$^{-1}$ and 718 W h kg$^{-1}$). As shown in Fig. 4d, the discharge curves of primary ZAB equipped with PC-Fe-50+RuO$_2$ cathode at different current densities (5, 10, 20 and 50 mA cm$^{-2}$) present voltage platforms of 1.24, 1.17, 1.08 and 0.8 V, remarkably higher than those of Pt/C+RuO$_2$ (1.19, 1.01, 0.6 and 0.27 V), indicating the outstanding discharge ability of PC-Fe-50. Fig. 4f presents the discharge polarization curves and the corresponding power density pots of the primary ZAB. The peak power density of the ZAB equipped with the PC-Fe-50+RuO$_2$ catalyst is calculated to be 128 mW cm$^{-2}$, higher than that of Pt/C+ RuO$_2$ (120 mW cm$^{-2}$). Furthermore, the cycling performance of the batteries was evaluated employing galvanostatic 10 min discharging and 10 min charging at 10 mA cm$^{-2}$ (Fig. 4g). Initially, the battery equipped with PC-Fe-50 catalyst affords a discharge voltage of 1.15 V and a voltage gap 0.87 V. After continuous charge/discharge cycles of 35h (around 105 cycles), the discharge voltage is 1.07 V with a small decay of 6% and the voltage gap increased to 1.08 V. In comparison, a deterioration of 27% is observed for the discharge voltage of Pt/C+RuO$_2$ air cathode in a shorter cycling period (60 cycles, around 20 h) and the voltage gap increased to 1.78 V, much bigger than that of PC-Fe-50+RuO$_2$. Moreover, the outstanding capacity, energy density and maxmum power density of ZAB equipped with PC-Fe-50+RuO$_2$ are comparable or better than of reported advanced catalysts (Table S7). These results sufficiently illustrate that PC-Fe-50 can be used as a promising substitute for commercial Pt catalyst applied in metal-air batteries.

## 3. Conclusion

In conclusion, we propose a facile and low-cost synthesis of single atomic FeN$_x$ sites anchored porous carbon catalyst with abundant N dopants. After optimizing the fabricating conditions, we obtained the PC-Fe-50 catalyst with impressive ORR activity superior to commercial 20% Pt/C and FePc. By analyzing the performance of PC-Fe-50 combining with DFT calculations, we propose the D2 structure combining the adjacent N atoms, as the realistic active sites because… (to be finished). The PC-Fe-50 was also successfully applied in rechargeable ZAB. Therefore, our work may offer a new strategy for the design and development of highly efficient and stable electrocatalysts for ORR as alternative for noble metal catalysts applied in electrochemical energy storage and conversion devices.